# ALD Zinc Tin Oxide Buffers for Chalcopyrite Solar Cells: Electrical Barriers and Conduction Band Cliffs

Boaz Koren[1,a)], Francesco Lodola[1], Zhuangyi Zhou[1], Tien Le[1], Kulwinder Kaur[1], Simon Backes[1], Michele Melchiorre[1] and Susanne Siebentritt[1,a)]

[1]Laboratory for Photovoltaics (LPV), Department of Physics and Materials Science,

University of Luxembourg, 41 rue du Brill, L-4422, Belvaux, Luxembourg.

a) Authors to whom correspondence should be addressed: boaz.koren@uni.lu and susanne.siebentritt@uni.lu

**ABSTRACT**

Sulfide chalcopyrite, Cu(In,Ga)S$_2$, having wide bandgap (>1.5 eV), favorable optoelectronic properties, and high stability, is a promising top-cell absorber for tandem applications. Adapting device structures optimized for 1.0–1.2 eV absorbers to wide bandgap absorbers requires modification of the buffer layer. This work investigates atomic layer deposition of ZnSnO as an alternative buffer layer to conventional CdS. A critical parameter for buffer performance is the conduction band offsets on both sides of the buffer. To investigate these buffers we electrically characterize solar cells utilizing different compositions of ZnSnO. The [Sn]/([Sn]+[Zn]) atomic ratio is controlled by the ratio of ZnO to SnO cycles during atomic layer deposition. Solar cells were fabricated utilizing CuInSe$_2$, Cu(In,Ga)Se$_2$, and Cu(In,Ga)S$_2$ absorbers, allowing cross-comparison with a variety of conduction band minimum energies. Buffer variation has two primary effects on cell performance: 1. Low tin buffers decrease the activation energy of interface recombination, reducing open circuit voltage. These observations indicates a cliff, a decrease of the conduction band minimum from absorber to buffer. 2. High tin buffers reduce the fill factor for all measured cells, and reduce the short circuit current under certain conditions. This observation indicates an electron transport barrier, conduction band offsets which limit the transport of electrons across the buffer, in either direction. We conclude that tin content correlates positively with the conduction band minimum of these buffers. Comparing different absorbers, cliffs occurs at lower Sn contents and the effects of barriers are more dramatic for absorbers with lower conduction band minima.

1. INTRODUCTION: ZTO BUFFERS FOR CHALCOPYRITE SOLAR CELLS

Tandem solar cells achieve higher power conversion efficiencies by integrating two cells with complementary bandgaps[1]. Chalcopyrites offer options for both the top and the bottom cells. A wide-bandgap sulfide chalcopyrite (~1.6 eV) absorbs high energy photons[2–5] while lower-energy photons are transmitted to the narrow-bandgap selenide chalcopyrite ($\leq 1$ eV) placed underneath[6–10]. Both of these cells require buffer layers to facilitate electron transport between the absorber and the window layers. Zinc tin oxide (ZTO) has been shown to be a viable buffer for both wide-bandgap sulfides $Cu(In,Ga)S_2$[2,11,12] and low bandgap selenides $Cu(In,Ga)Se_2$[13–16]. While the commonly used CdS buffer is suitable for low bandgap chalcopyrites and has been used in the record chalcopyrite solar cell[17], CdS shows an unfavorable band alignment with wide-bandgap chalcopyrites[2,18] and alternative buffer layers are needed for high efficiency.[19]

This study investigates atomic layer deposition (ALD) of different ZTO buffer layers on various chalcopyrite absorbers. In addition to sulfur based solar cells with a bandgap of 1.6 eV, selenium based absorbers with bandgaps of 1.14 eV and 1.0 eV are fabricated and analyzed. ALD ZTO has multiple properties making it desirable as a buffer layer: ZTO has a wide-bandgap, $>3$ eV[20–22], and thus minimal parasitic absorption, increasing output current. ALD allows fine thickness control and high conformity[23] while also enabling control over material composition. In principle, by tuning the composition, the bandgap and the band offsets to adjacent layers can be adjusted.

The main control parameter of ZTO which is adjusted in this study is the atomic ratio: [Sn]/([Sn]+[Zn]) (TTZ). Since the buffer constitutes the electron contact to the absorber, the conduction band minimum (CBM) is highly relevant for solar cell efficiency. Present literature[15,20,24] is not in agreement on the effects of changing TTZ on ZTO properties. The studies contradict each other on both the direction and amount of CBM energy – TTZ dependence. While the ZTO buffers are defined in this study mainly by their composition, studies have shown that deposition parameters such as supercycle size[20] and substrate temperature[16,22] can change the ZTO's bandgap by hundreds of meV, without significantly altering composition. This suggests that material properties other than composition also affect buffer performance, likely properties such as density or secondary phases[22]. Even though we do not investigate these parameters which could vary in other systems, we expect the general insights and trends to extend beyond our specific process.

Absorber CBM is also dependent on composition, of the absorbers investigated in this study: CuInSe$_2$ has the lowest CBM, Cu(In,Ga)Se$_2$ exhibits a slightly higher CBM, and the sulfides' CBMs are higher still[25], though the exact differences are difficult to determine. Furthermore, the Ga containing absorbers have a Ga gradient[2,18,26] and the exact levels of Ga (and thus CBM) at the absorber-buffer interfaces are unknown.

CBM alignment between the buffer and its adjacent layers is critical for solar cell efficiency. Sub-optimal CBM alignment can affect cell performance in multiple ways [27–30] (Fig. 1):

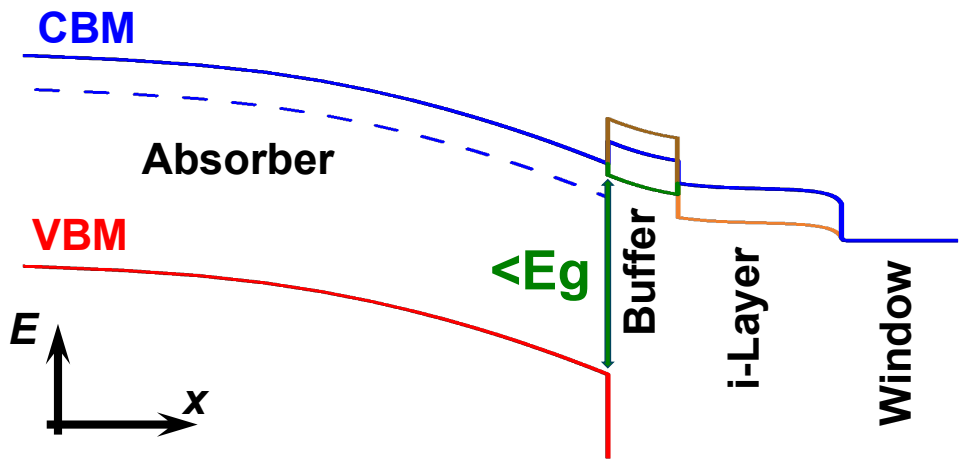

FIG. 1: Sketch of conduction band alignment between the absorber, buffer, i-layer and window. The solid blue profile represents favorable alignment, small spike from absorber to buffer, small cliff from buffer to i-layer. The green profile increases interface recombination due to a cliff at the absorber-buffer interface, and brown and/or orange profiles create electron transport barriers at the absorber-buffer interface and/or the buffer-i-layer interface. The dashed blue line describes the CBM of an absorber a with narrower band gap and lower CBM.

If the buffer CBM is lower than the absorber's, the negative conduction band offset (CBO) is called a cliff. The result is an interface where the difference between the buffer's CBM and the absorber valence band maximum (VBM) is smaller than the absorber's bandgap. This increases non-radiative recombination, reducing open circuit voltage ($V_{OC}$), and fill factor ($FF$)[27,28]. If the buffer's CBM is higher than the absorber's, the positive CBO is called a spike, and if it is too large, it can form a barrier for the photocurrent. A large cliff at the buffer-i-layer interface can also present a barrier, limiting the forward diode current. Both barriers affect the $FF$; a large barrier for the photocurrent significantly affects short circuit current density ($J_{SC}$) while a barrier for the forward current significantly affects the current near and beyond $V_{OC}$[16,27,28]. For optimal performance, the desired alignment is a small spike from absorber to buffer. A cliff between buffer and i-layer is acceptable, as long as it is not too high. The exact CBO size that significantly impacts the current depends on multiple factors, such as the amount of photocurrent and thus the absorber bandgap, as well as the changes in electrostatic potential across the buffer and i-layer. According to Sood et al.[27], for a sulfide absorber with a bandgap of ~1.6 eV, an absorber-buffer spike larger than 350 meV or a buffer-i-layer cliff higher than 300 meV become detrimental.

In order to investigate the performance of ZTO buffers, we measure absolute photoluminescence (PL) to determine quasi-Fermi level splitting in the absorber, JV (current density-voltage) characteristics to analyze diode and photovoltaic parameters, JVT (temperature dependent JV) characteristics to determine activation energies, and EQE (external quantum efficiency) spectra to learn about photocurrent generation. We use an array of 8 buffers on 4 absorbers: two sulfide absorbers, Cu(In,Ga)S$_2$, (labelled CIGSu for clarity) and two selenide absorbers, with Ga (Cu(In,Ga)Se$_2$, labelled CIGSe) and without Ga (CuInSe$_2$, labelled CISe). The four absorbers are identified by their composition, back contact, and bandgap (derived from EQE): 1. CIGSu-Mo, 1.60 eV, 2. CIGSu-ITO, 1.61 eV, 3. CIGSe (Mo), 1.14 eV, 4. CISe (Mo), 1.00 eV. The indium tin oxide (ITO)'s transparency is critical for application as top-cells[3] in tandem structures. We include also CIGSu on (opaque) Mo back contact, because the absorber deposition process was originally developed on Mo. In accordance with previously reported results, the selenides employ ZnO i-layers, and the sulfides employ Al:ZnMgO i-layers[2,18,27]. Further details are provided in the "device preparation" section. We utilize 8 buffers, one chemical bath deposited (CBD) CdS, and 7 ALD ZTO, where zinc oxide and tin oxide ALD cycle ratios are varied. The resulting compositions, measured by energy-dispersive X-ray spectroscopy (EDX) on Si witness samples, are TTZ of 0 (ZnO), 4, 9, 19, 20, 28.5 and 38%. This set of 30 cells (Two of the CIGSu-ITO cells were damaged during preparation: 4% and 19% TTZ) utilizing different absorbers allows us to analyze trends in solar cell parameters with respect to buffer TTZ.

## II. RESULTS AND DISCUSSION

The dependence of solar cell performance on buffer composition are given in Fig. 2 for CIGSu-Mo and CISe. We can already observe some shared trends: Low TTZ leads to losses in $V_{OC}$, high

TTZ to losses in $FF$, and to $J_{SC}$ reduction in the selenides. Both trends are compatible with the assumption that higher TTZ leads to a higher CBM in the ZTO buffers: low TTZ leads to a cliff at the absorber-buffer interface and reduces $V_{OC}$, high TTZ creates a large spike at the absorber-buffer interface and a high cliff at the buffer-i-layer interface, creating barriers for photo and forward current, respectively, reducing $FF$ and potentially $J_{SC}$. In the following we discuss these trends in detail.

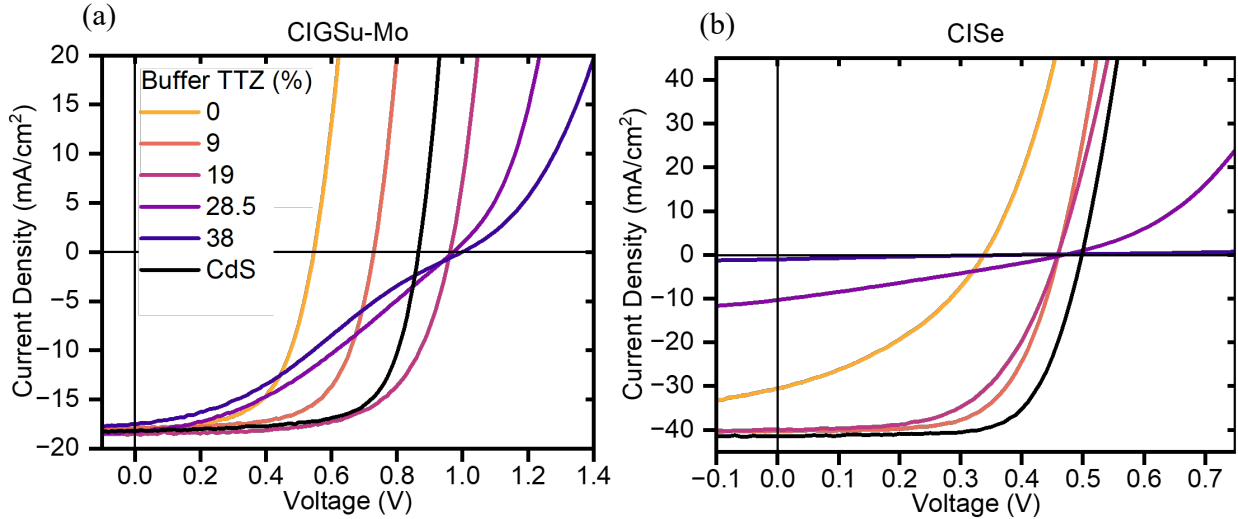

FIG. 2: JV plots for representative CIGSu-Mo (a) and CISe (b) cells, as tin content increase, $V_{OC}$ increases, $FF$ decreases and for the selenides, $J_{SC}$ decreases. The legend in (a) is valid for (b). Note the different scales, to accommodate the higher voltage of the sulfides and the higher current of the selenides.

A. QFLS, $V_{oc}$ and Activation Energies

Reductions in $V_{OC}$ of chalcogenide solar cells are primarily caused by non-radiative losses, either in the absorber bulk in the area of the absorber-buffer interface[31,32]. To distinguish between the two effects we perform absolute PL (Fig. 3a), to determine quasi fermi level splitting (QFLS)[32,33] of the bare absorbers and the absorbers with buffers[32]. In contrast to sulfide absorbers, bare selenides degrade too quickly in air[34–36] and are thus only measured with buffers. Due to our method of analysis, QFLSs of the sulfides are underestimated by about 20meV and of selenides by about 5meV. Further details are provided in the "cell characterization" section. Our first observation (Fig. 3) is that the QFLSs of the sulfide absorbers with buffers are lower than those of the bare absorbers. These losses are quite large: they correspond to a decrease in PL intensity of about an order of magnitude,[32] which cannot be solely explained by changes in reflectivity. We must therefore conclude that all buffers increase interface recombination in these sulfide absorbers. The following rough trend is observed for all absorbers: QFLS is generally lower for low TTZ, but levels to a mostly constant value at higher TTZ. The buffer composition at which the levelling off occurs is higher for sulfides (around 20% TTZ) than for selenides (5-10 % TTZ). A similar, but stronger trend is observed for $V_{OC}$ of the completed solar cells (Fig. 3b).

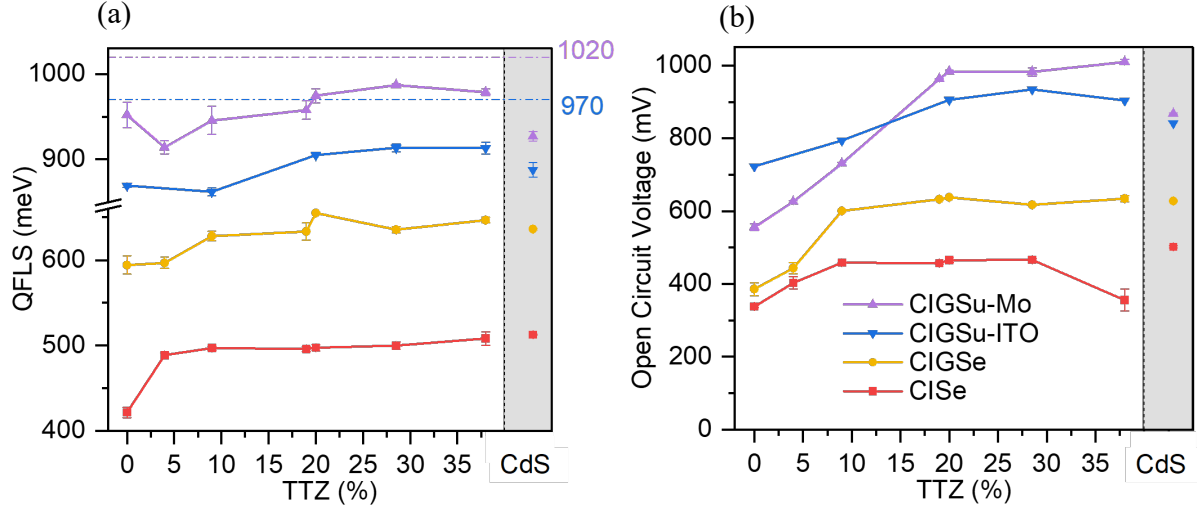

**FIG. 3:** (a) Quasi fermi level splitting after buffer deposition, the dash-dotted lines marks QFLS for the bare sulfides. (b) Open circuit voltage. The legend in (b) is valid for (a).

Comparing now the $V_{OC}$ dependencies on TTZ for the different absorbers: Starting at 0%, as TTZ increases, $V_{OC}$ increases, until a plateau is reached at a certain buffer composition. This composition is different for the different absorbers and corresponds to the bandgap (and CBM energy[25] – compare the solid and dashed absorber CBM lines in Fig. 1): ~9% for the selenides and ~20% for the sulfides. While both selenide series show significant $V_{OC}$ reductions below 9% TTZ, the considerably larger decrease between 9 and 4% for CIGSe implies a larger cliff than in CISe. The difference between $V_{OC}$ and QFLS (Fig. S1, supplementary material) is indicative of interface recombination causing a gradient in the electron quasi Fermi level[31]: increased interface recombination slightly reduces QFLS in the bulk, but, with moderate mobilities, has a significantly larger impact at the interface. Absolute PL essentially measures the highest QFLS, whereas $V_{OC}$ depends on the electron Fermi level at the electron contact[32]. Therefore the large cliffs at low TTZ, which enhance interface recombination, have a much larger impact on $V_{oc}$ than on QFLS. The CdS buffer's $V_{OC}$ falls somewhere between those corresponding to TTZs of 9 and 19% for all but the CISe, where it is higher than all ZTOs. The reduced $V_{OC}$ of the CISe-38% TTZ cell is due to the fact that the cell is almost completely blocked (Fig. 2b).

We additionally measure temperature dependent JV for select cells at a temperature range of 80-300K (Fig. 4a). Temperature reduction increases $V_{OC}$ while reducing *FF* and $J_{SC}$. The dependence of $V_{OC}$ on temperature can be derived from the diode model[28,37]:

$$V_{oc} \approx \frac{E_A}{q} - \frac{nk_BT}{q}\ln\left(\frac{J_{00}}{J_{SC}}\right) \tag{1}$$

$n$: diode ideality factor; $k_B$: Boltzmann constant; $T$: temperature; $q$: unit charge; $E_A$: activation energy of the dominant recombination process; $J_{00}$: reference current density.

We assume that $J_{00}$ and $n$ have only weak temperature dependencies. Eq. 1 then implies that $E_A$ can be derived through a linear extrapolation of $V_{OC}(T)$ to 0K (Fig. 4b). We assume that most of the recombination happens either in the bulk (in which case $E_A$ = bandgap) or near the absorber-buffer interface. In cases where there is a cliff at the absorber-buffer interface, the activation energy of interface recombination is smaller than the absorber's bandgap. The dominant recombination process is usually the one with the lowest energy. Analyzing the activation energies as a function of buffer composition (Fig. 4c), we see that the trends are similar to those found in room temperature $V_{OC}$ analysis. $V_{OC}$ losses at low TTZ are therefore understood to be a result of additional interface recombination due to CBM cliffs at the absorber-buffer interface. Since the sulfide's $V_{OC}$ increasing with TTZ up to 20%, we conclude that at least in that composition range, TTZ is positively correlated with buffer CBM.

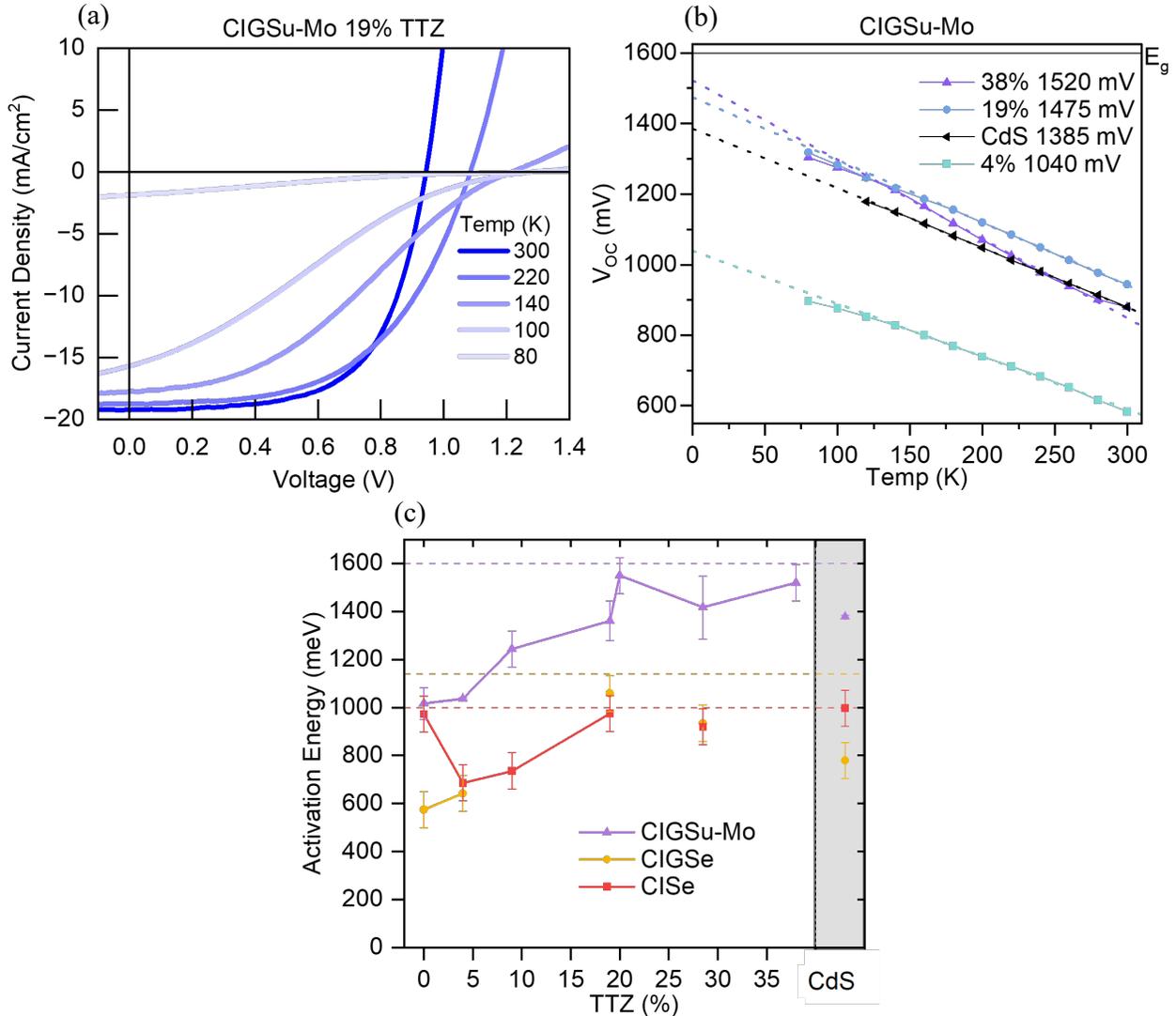

FIG. 4: Temperature dependent of JV characteristics under illumination (a) JV characteristics of CIGSu-Mo 19%TTZ at different temperatures (b) $V_{OC}(T)$ plot of CIGSu-Mo cells with extrapolations to 0K, corresponding to the activation energy of the reverse saturation current (c) JVT derived activation energies, bandgaps are indicated by dashed lines (for samples with only one measurement, the error is estimated at 75meV).

B. Fill Factor and Electron Transport Barriers

The *FF* trends (Fig. 5a) form somewhat of a mirror image of the $V_{OC}$ trends: The *FF* is higher with TTZ ≤19% buffers, and present a sharp decrease from 19 to 20% TTZ. This is considerably less sudden if the composition x-axis is replaced with the ALD cycle ratios (Fig. S2). This trend is observed in all absorbers besides Cu(In,Ga)S$_2$-ITO, where the *FF* is rather low for all buffers due to the high series resistance caused by the transparent back contact[3]. We focus our analysis of *FF* on the CIGSu-Mo cells: To better see the changes in the shape of the JV characteristics, we shift the voltage from Fig. 2a by $V_{OC}$ (Fig. 5b). We see that the shape of the JV curve around $V_{OC}$ hardly changes from 0 to 9% TTZ. This suggests that indeed the activation energy is main parameter which varies in this range, as opposed to the resistances or diode factor. As TTZ continues to increase, we see mild changes at 19% TTZ, and at higher values a clear "S-shape" emerges. This indicates a barrier, most likely for the photocurrent, as there is no roll-over above $V_{OC}$. Cells with CdS buffers result in superior *FF* for all absorbers, we attribute this advantage not to the CBOs, but rather to the chemical bath deposition process. The ammonia involved in CdS deposition removes oxides and other impurities from the absorber-buffer interface, likely reducing surface defects. As selenides oxidize considerably more than sulfides[36,38], the potential oxide removal in CBD is much more significant, leading to a larger difference between the *FF* with CdS and the best *FF* with ZTO buffer. Previous experiments in our lab with similar selenide absorbers and ALD ZTO buffers (Fig. S3) have resulted in considerably higher *FF*s, indicating that the low selenide-ZTO *FF*s presented in Fig. 5 are avoidable. The lower *FF*s of selenides in the series here are most likely caused by re-oxidation of the absorber surface, due to the rather long time between cleaning by KCN etch and buffer deposition.

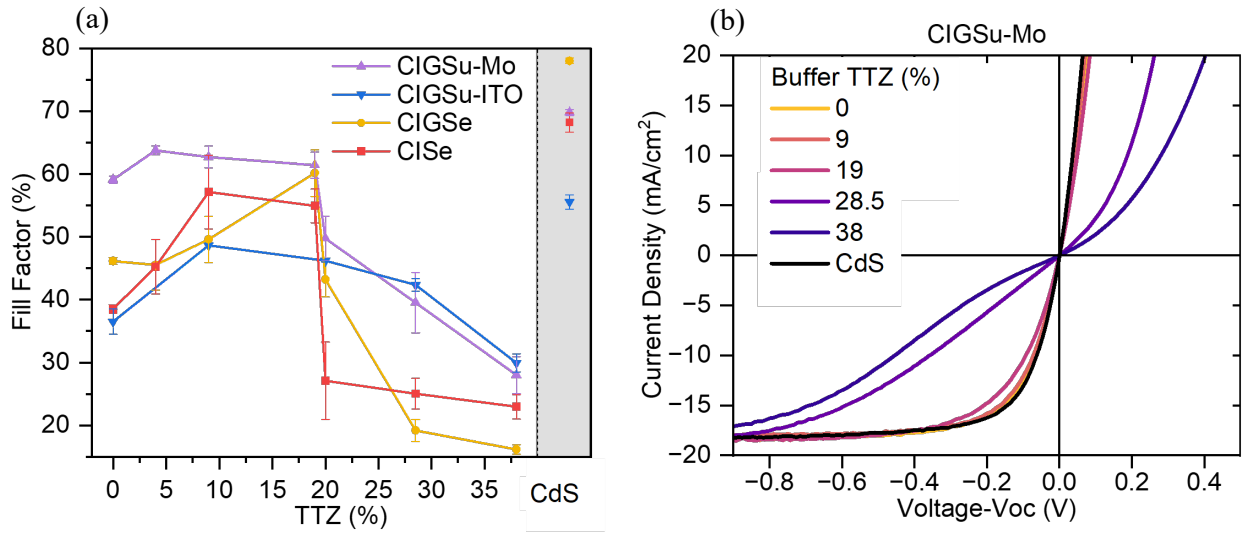

FIG. 5: (a) Fill factor with TTZ variation for different cells. (b) JV of selected CIGSu-Mo cells, centered around $V_{OC}$.

To describe the trends observed in *FF* and $J_{SC}$, we remind that large barriers limit current transport exponentially[27,28,39]. The trends for barriers are described here in terms of CBOs: large spikes block the photogenerated current, large cliffs block the forward current. Electron transport barriers are defined as the difference between the electron quasi fermi level and the highest CBM near the interface, but when analyzing comparatively, CBO examination is sufficient. Both barriers are expected to be larger for selenides than for sulfides. The lower bandgap (and CBM[25]) results in larger spikes at the absorber-buffer interface and the lower CBM i-layer[27,40] utilized for the selenides results in larger cliffs at the buffer-i-layer interface.

The decrease in *FF* for high TTZ is attributed to a high CBM in ZTO with high Sn content. It forms a barrier for the photocurrent and likely also for the forward current[27]. We do not observe a roll-over in the forward voltage at room temperatures, but it is observed at lower temperatures (Fig. 4a). The *FF* decrease at 20% TTZ and higher is stronger for the selenide cells than for the sulfide cells, as expected (compare the solid and the dashed blue line in Fig. 1). We therefore conclude that the ZTO CBM correlates positively with the TTZ, also in the 20-38% range.

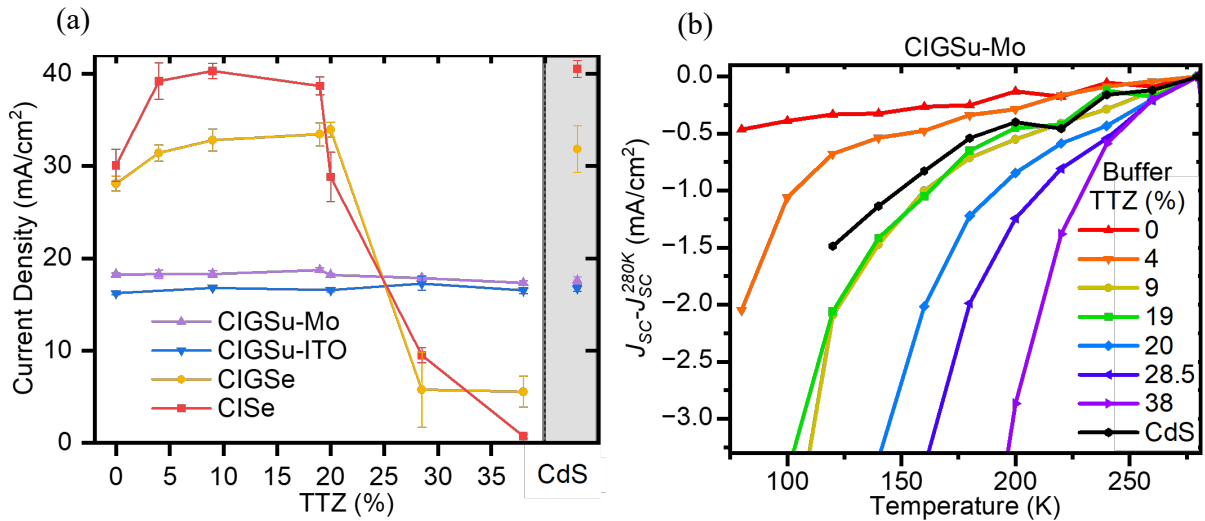

**FIG. 6: (a) Short circuit current density of the cells (b) $J_{SC}$ of CIGSu-Mo cells as a function of temperature, relative to $J_{SC}$ at 280K.**

C. Short Circuit Current and External Quantum Efficiency

From the data shown in Fig. 6a, it is evident that $J_{SC}$ of the sulfides is buffer independent, while $J_{SC}$ for the selenides is greatly reduced by buffers with TTZ above 20%. In these cases the barrier for the photocurrent is so high that it impedes the short circuit current. This effect is more pronounced in the low CBM selenides than the high CBM sulfides (dashed versus solid blue lines in Fig. 1). There is also a small reduction of $J_{SC}$ for selenides with pure ZnO buffers which may be due to a very defective interface, that negatively alters the whole device band structure. Investigation of the sulfides at lower temperatures revealed that cells with higher TTZ buffer also suffer reductions of $J_{SC}$ (Fig. 6b) at lower temperatures, more so the higher the TTZ (and CBM). These results support the model of a barrier for the photogenerated electrons at the absorber-buffer interface. Additional electrical characterizations can be found in Fig. S4.

III. SUMMARY AND CONCLUSIONS

A wide study consisting of matching different chalcopyrite absorbers with different ALD ZTO buffers has been conducted, striving to find trends in ZTO CBM energy relevant for solar cells and endeavoring to identify the various losses in efficiency caused by different buffers. Our conclusions are that the CBM increases with TTZ from 0 to 38%, based on observations of the three parameters determining efficiency: $V_{OC}$, $FF$ and $J_{SC}$: Buffer dependent $V_{OC}$ losses are attributed to CBM cliffs between absorber and buffer, and we see lower $V_{OC}$ for low TTZ buffers, especially those on high CBM absorbers (sulfides). The reduction in the activation energy of the reverse saturation current for low TTZ buffers, observed from temperature dependent JV measurements confirms our conclusion of a cliff increasing with lower TTZ. $FF$ losses are expected when electron transport across the buffer is hampered, either by a large CBM spike at the absorber-buffer interface, or by a large CBM cliff at the buffer-i-layer interface. We record significant reductions of $FF$ for high TTZ buffers, particularly for cells with low CBM absorbers (selenides) and i-layers (ZnO). $J_{SC}$ losses are expected to correlate with large spikes between absorber and buffer, and we see significant losses with the high TTZ buffers at room temperatures for the selenides and at low temperatures for the sulfides.

This control over buffer CBM, combined with ZTO's wide bandgap (>3 eV) and low toxicity, as well as the various advantages of ALD such as thickness control, uniform coverage, and relatively low operation temperatures (120°C) makes ZTO a promising buffer material for chalcopyrites.

IV. EXPERIMENTAL SECTION

1. DEVICE PREPARATION

A. Solar Cell Structure

The general solar cell structure is: glass-back contact-absorber-buffer-i-layer-window-grids (Fig. S5). The back contact is Mo for the CIGSu-Mo and the selenides and ITO for the CIGSu-ITO cells. The Ga containing absorbers are prepared with a Ga gradient, increasing towards the back contact to decrease backside recombination[26]. To reduce backside recombination in the CISe, a Ga oxide based hole transport layer is employed [10], as in Wang et al.[41].

The absorbers are Cu(In,Ga)S$_2$, Cu(In,Ga)Se$_2$ and CuInSe$_2$. The buffer layers are systematically varied throughout this study. The i-layer is ZnO for the selenides and Al:ZnMgO for the sulfides. In all cases the window is Al:ZnO and the grids are made from Ni and Al. Further details are provided in Table S1.

B. Absorbers

The four absorbers were deposited using three-stage co-evaporation[18,42]: 1. CIGSu-Mo, 1.60 eV, 3 μm. Silver and sodium fluoride are added to increase grain size, doping and $V_{OC}$[5,43]. 2. CIGSu-ITO, 1.61 eV, 1.5 μm. The transparent ITO is critical for tandem applications.[1,3] 3. CIGSe, 1.14 eV, 2.5 μm. 4. CISe, 1.00 eV, 2.2 μm. Absorber bandgaps are identified by the inflection point of the EQE spectrum.

The selenides received a 5% potassium cyanide (KCN) etching for 30 seconds one day before buffer deposition, and were then stored in evacuated bags until ALD deposition, the sulfides received no treatment.

C. ZTO and CdS Buffers

The ALD ZTO layers are deposited in a BENEQ TFS-200 platform, the precursors used are diethyl zinc (DEZn) and tetrakis-dimethyl-amino tin (TDMASn) with deionized water (H$_2$O) as a co-reactant and nitrogen (N$_2$) as an inert carrier. We deposit ZTO with cycles of either DEZn-N$_2$-H$_2$O-N$_2$, or TDMASn-N$_2$-H$_2$O-N$_2$. We use the following exposure/purge times (in seconds): 0.5-2-0.3-2 for zinc oxide, and 0.6-2-0.5-2 for tin oxide, and a deposition chamber temperature of 120°C. Buffer composition is controlled by the zinc to tin cycle ratio, where minimally sized "supercycles" are employed to reduce layering effects[20]. Si wafers placed next to the samples are used for composition measurement by 7 KV EDX and thickness by ellipsometry. The growth statistics for the buffers used in this study, as well as previous depositions in our lab with the similar parameters are presented in Fig. S6. We suspect that tin content may be overestimated in this study, samples measured by EDX at 21 and 30% TTZ were measured in a Rutherford back scattering system at 17 and 25% TTZ respectively.

The CdS chemical bath is composed of 2 mM CdSO$_4$, 50 mM of thiourea, and 1M of ammonia, and deposition takes place at 67 °C for ~7 minutes[41].

## 2. Cell Characterization

### A. JV, JVT and EQE

JV measurements are performed in a AAA solar simulator, calibrated with a reference standard Si solar cell. The data shown in this study is composed of statistics on 4 sub-cells for each absorber-buffer pair, with some missing due to shunts. A second JV setup with a cryostat is used for JVT using a halogen lamp, illumination level is calibrated by matching $J_{SC}$ to the one measured under the solar simulator. EQE was measured utilizing a lock-in amplifier and using chopped illumination with two lamps, halogen for low wavelengths and Xenon for high, switched at 850 nm. Light intensity is measured by two commercial photodetectors, Si for low wavelengths and InGaAs for high, switched at 960nm.

### B. Photoluminescence

Absolute PL measurements were conducted using an in-house setup with both spectral and intensity calibration[33]. Samples were excited using a continuous-wave laser, 637nm wavelength for selenides, 450 nm for sulfides with a photon flux equivalent to the flux of the AM1.5G spectrum above the absorber's bandgap. For selenides, QFLSs were extracted by fitting the high-energy tail of the PL spectrum, assuming an absorptance of 1 (this assumption may lead to a slight underestimation of about 5 meV in QFLS, as A(E) < 1 in this region)[44] using Planck's generalized law under the Boltzmann approximation. The fit is performed with a fixed temperature of 296 K, as measured in the lab. For the sulfides, fitting the high-energy wing of the PL spectrum is not feasible, as fixing the temperature leads to a completely inaccurate fit. In this case, QFLS is extracted using the relation[32]:

$$\text{QFLS} = V_{OC}^{SQ} + k_B T \ln(\text{PLQY}) \qquad (2)$$

Where the PL quantum yield (PLQY) is calculated by integrating the PL spectra, without considering the luminescence of defects. To calculate the $V_{OC}^{SQ}$, the PL peak was used to estimate the bandgap. This approach potentially resulted in an underestimation of the QFLS by approximately 20 meV.[33]

SUPPLEMENTARY MATERIAL

The supplementary material includes diagrams, tables, and results supporting the arguments made in this study.

ACKNOWLEDGMENTS

The authors acknowledge the support from the European Union within the SITA project (no. 101075626).

We thank doctor Tobias Törndahl for the university of Uppsala for RBS calibration measurements, and many discussions regarding the ALD process.

We thank professor Nicolas Barreau from University of Nantes for top contact depositions for the selenides.

We thank our lab engineer Thomas Schuler for all kinds of technical support.

AUTHOR DECLARATIONS

**Conflict of Interest**

The authors have no conflicts to disclose.

**Author Contributions (CRediT)**

**Boaz Koren**: Conceptualization, Data curation, Formal analysis, Investigation, Methodology, Writing. **Francesco Lodola:** Investigation, Resources, Writing. **Zhuangyi Zhang:** Resources, Software. **Tien Le:** Investigation, Resources. **Kulwinder Kaur:** Resources, Writing. **Simon Backes:** Investigation. **Michele Melchiorre:** Resources. **Susanne Siebentritt**: Conceptualization, Funding acquisition, Methodology, Project administration, Supervision, Writing.

## Data Availability

The data analyzed in this study will be openly available on Zenodo, will finalize pending acceptance.

# Supplementary Material

## ALD Zinc Tin Oxide Buffers for Chalcopyrite Solar Cells: Electrical Barriers and Conduction Band Cliffs

Boaz Koren[1,a], Francesco Lodola[1], Zhuangyi Zhou[1], Tien Le[1], Kulwinder Kaur[1], Simon Backes[1], Michele Melchiorre[1] and Susanne Siebentritt[1,a]

[1]Laboratory for Photovoltaics (LPV), Department of Physics and Materials Science,

University of Luxembourg, 41 rue du Brill, L-4422, Belvaux, Luxembourg.

a) Authors to whom correspondence should be addressed: boaz.koren@uni.lu and susanne.siebentritt@uni.lu

S1: *Voc*-QFLS: A large difference between q$V_{OC}$ and QFLS indicates strong interface recombination[1,2]. The difference is therefore labelled interface loss. A few values are higher than zero. This is due to the underestimation of QFLS in sulfides by about 20meV (see methods section). It is interesting to look first at the values for the CdS buffers. The selenides show little interface loss, as observed previously and is expected since CdS buffers are optimized for selenides. The interface loss in the sulfides with CdS is considerably larger, indicting the need for an alternative buffer for these wide gap absorbers. In the high TTZ range the interface loss is almost constant and quite small (besides the CISe-38% cell, which is almost completely blocked – see Fig. 2b). For TTZ lower than 19% for sulfides and lower than 9% for selenides the interface loss increases, again confirming the dominance of interface recombination, caused by a cliff type band alignment in that composition range. The exact amount of interface loss depends on many factors[1], like band offset, defect density, carrier mobility and absorber thickness.

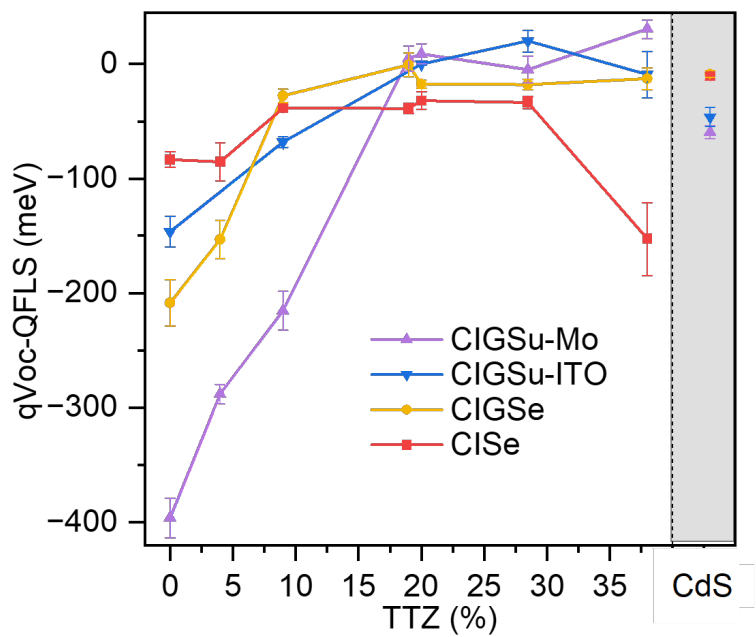

**FIG. S1: Difference between $V_{OC}$ (Fig. 3b) and QFLS (Fig. 3a)**

S2: Fig. 5a with a recalibrated x-axis representing the deposition parameters rather than the resulting composition. This figure illustrates that buffer properties are not only a product of composition, but other material properties as well. These other properties are perhaps more affected by the deposition parameters than the composition is.

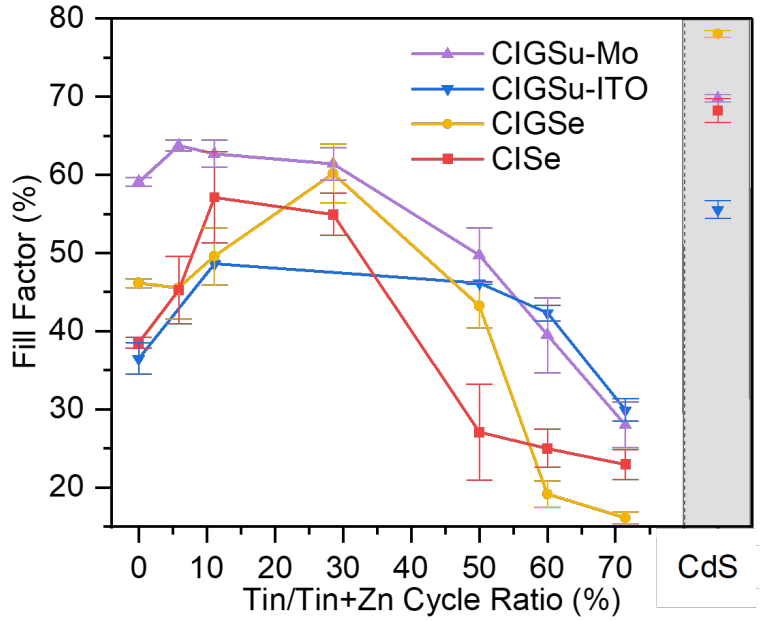

**FIG. S2: Fill factor as a function of Zn/Zn+Sn cycle ratio during buffer deposition.**

S3: Previous good selenides: JV characterization parameters from a previous experiment using cells similar to the selenides analyzed in this study, ~1.5-2μm thickness, CGI of 75% for CISe B and 85%-90% for the others, GGI of 21% for CIGSe A, and 28% for CIGSe B. Several of the ZTO buffers result in *FF* of over 70%, showcasing that our process is capable of producing selenides with high *FF*.

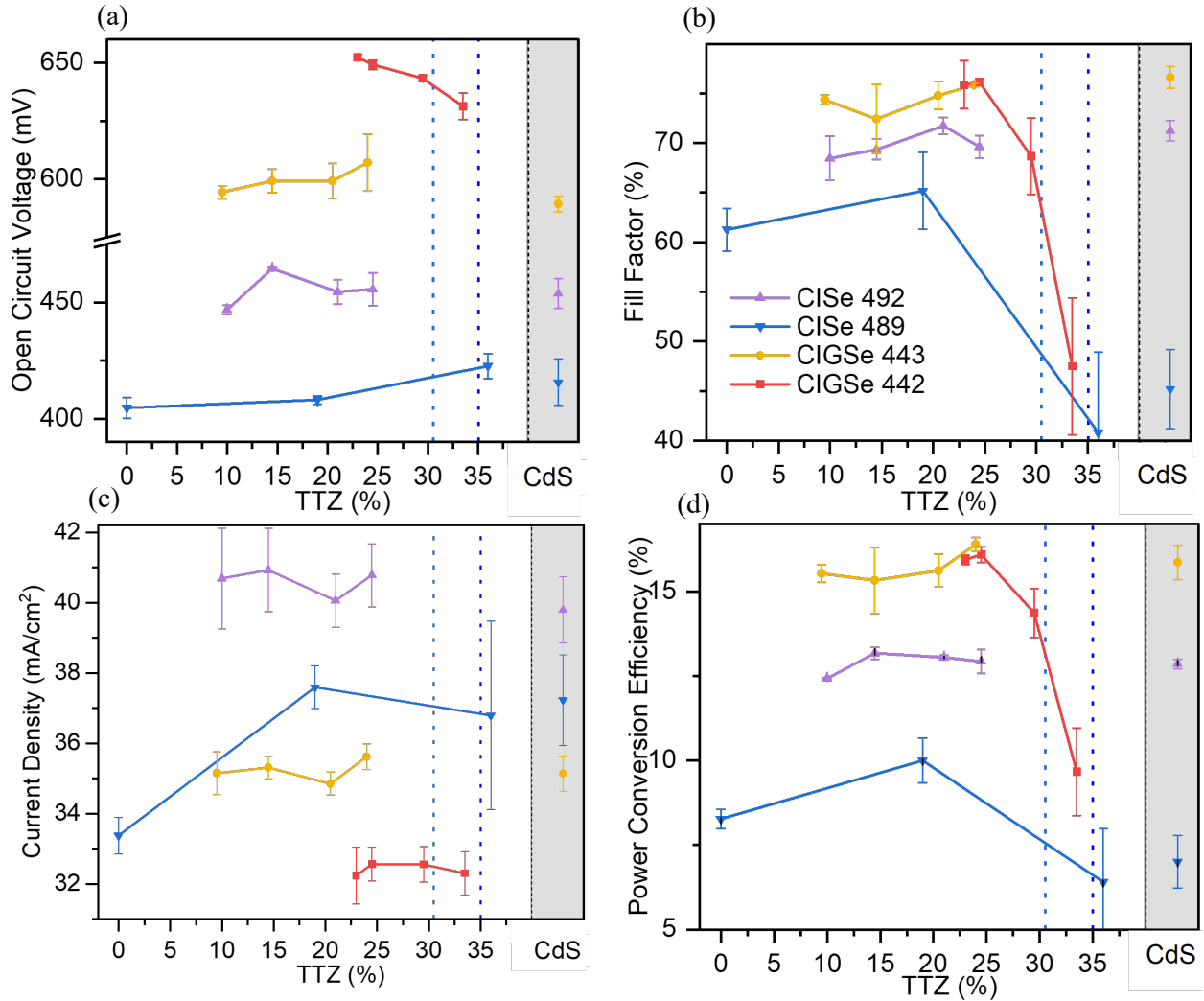

**FIG. S3: JV results from previous experiment: The dotted blue lines mark two CISe B cells where the current was severely blocked (a) Short circuit current (b) Fill factor (c) Open circuit voltage (d) Power conversion efficiency. The legend in (b) is valid for all.**

S4: Extra graphs: Additional figures for EQE and PCE, and extra graphs for selenides. The EQE for all ZTO sulfides is quite similar. The CdS buffers cause a small reduction in EQE for wavelengths below 500nm (CdS bandgap wavelength is 520nm). The trends exhibited in the CIGSe cells do not really add much beyond what can be gleaned from the CIGSu-Mo and CISe analyses.

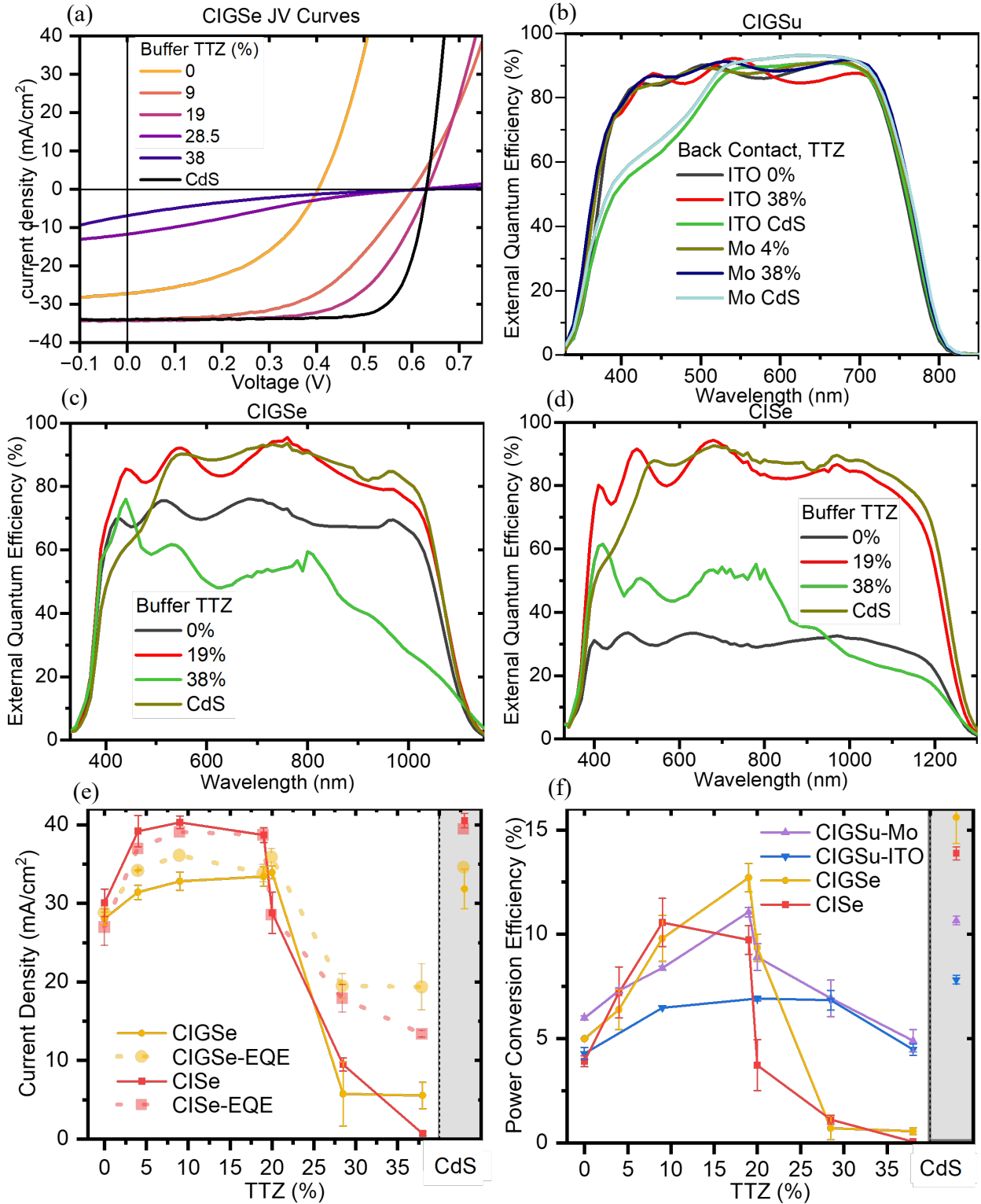

**FIG. S4: (a)** JV of selected CIGSe cells. **(b-d)** EQE of the cells, **(e)** Comparison of $J_{SC}$ measured in JV and calculated from EQE, **(f)** Power Conversion Efficiency.

S5: Extra deposition details: a diagram of the CIGSu-Mo cells structure (Fig. S5) and a table (table S1) providing more detail regarding all of the cells. Each cell is based on optimal cell (at the time) production by different lab members.

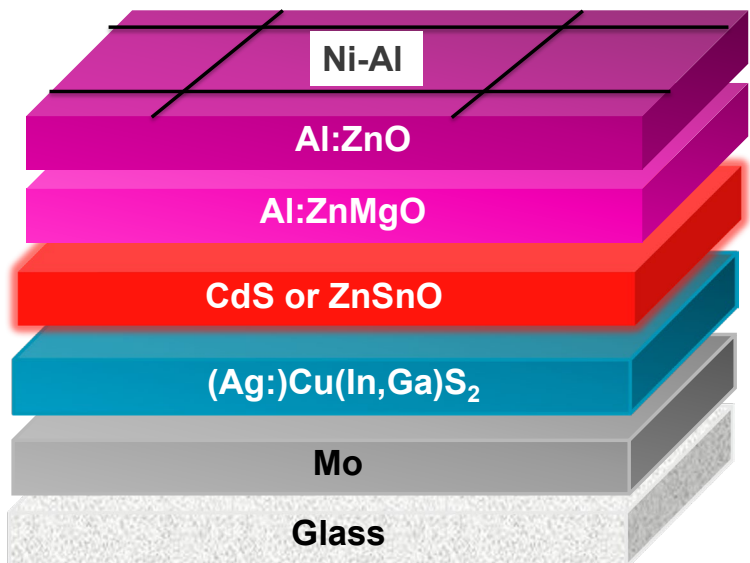

**FIG. S5: Diagram of the CIGSu-Mo cells.**

| Layer\Cell | CISe | CIGSe | CIGSu-ITO | CIGSu-Mo |
|---|---|---|---|---|
| Top contact (E-beam) | 30nm Ni+2.5μm Al+30nm Ni | 30nm Ni+2.5μm Al+30nm Ni | 10nm Ni+3μm Al | 10nm Ni+3μm Al |
| Window (Sputter) | 380nm 2%Al doped ZnO | 380nm 2%Al doped ZnO | 380nm 2%Al doped ZnO | 380nm 2%Al doped ZnO |
| i-layer (Sputter) | 80nm ZnO | 80nm ZnO | 64 nm Al:ZnMgO~26% [Mg]/[Mg+Zn] and ~0.5% Al | 64 nm Al:ZnMgO~26% [Mg]/[Mg+Zn] and ~0.5% Al |
| Buffer (ALD/CBD) | ~50nm ZTO or CdS | ~50nm ZTO or CdS | ~50nm ZTO or CdS | ~50nm ZTO or CdS |
| Absorber (PVD) | CISe, 1eV, 2.2μm | CIGSe, 1.14 eV, 2.5μm | CIGSu, 1.61eV, 1.5μm | CIGSu, 1.6eV, 3μm, alloyed with ~3% Ag and Na |
| groupI/groupIII atomic ratio (20KV EDX) | 95% | 87% | 92% | 95% |
| GGI (20KV EDX) | 0% | 21% | 19% | 25% |
| Hole transport layer (Sputter) | 120nm CISe covered by 30nm GaO$_X$ | - | - | - |
| Back contact (Sputter) | 500nm Mo | 500nm Mo | 100nm ITO | 500nm Mo |
| Glass Substrate thickness (mm) | 2 | 2 | 1.1 | 2 |

**Table S1: Cells layer details.**

S6: ALD depositions: details for the depositions of the buffers analyzed in this work and also previous depositions with the same parameters, varying only cycle ratio. Each datapoint in the graph is one deposition, with statistics for 6 EDX measurements at different positions around the chamber (avoiding the gas inlet, where deposition is unstable). The line y=x has been added to Fig. S6a as a visual aid. The ALD depositions resulting in 20 and 28.5% TTZ had slightly lower TTZ than expected. The 9%TTZ deposition was slightly thinner than expected at ~43nm, while the second thinnest ZTO was ~47.5nm. The details of these depositions, including EDX and ellipsometry results, are presented in table S2:

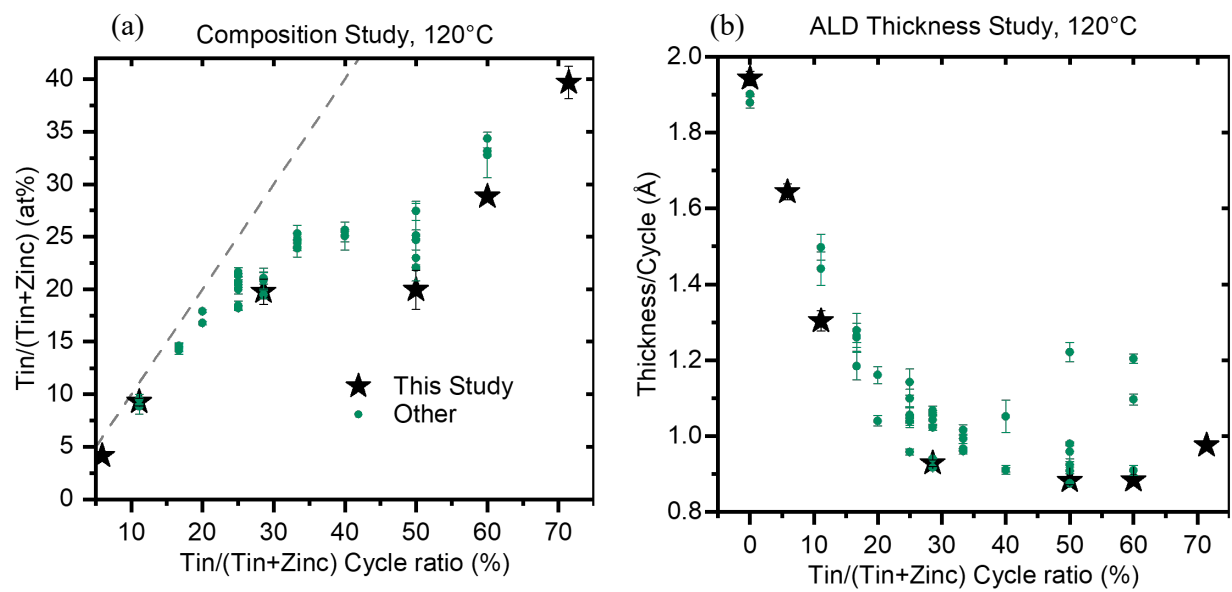

**FIG. S6:** ALD calibration: (a) composition analysis, the dashed line marks y=x (b) deposition rate analysis. The legend in (a) is valid for (b).

| Cycle Pattern (Zn:Sn:Zn:Sn) | 1:0 | 16:1 | 8:1 | 3:1:2:1 | 1:1 | 1:2:1:1 | 1:3:1:2 |
|---|---|---|---|---|---|---|---|
| Cycle ratio [%] | 0 | 5.9 | 11.1 | 28.6 | 50 | 60 | 71.4 |
| TTZ (7KV EDX) [%] | 0 | 4 | 9 | 19 | 20 | 28.5 | 38 |
| Number of Cycles | 263 | 289 | 333 | 546 | 560 | 550 | 511 |
| Thickness (ellipsometer) [nm] | 51.1 | 47.5 | 43.4 | 50.7 | 49.4 | 48.6 | 49.8 |

**Table S2:** ALD depositions.